\documentstyle[12pt]{article}
\newcommand{\dd}{{\rm d}}
\newcommand{\p}{\partial}
\newcommand{\nnn}{\noindent}
\newcommand{\be}{\begin{equation}}
\newcommand{\ee}{\end{equation}}
\newcommand{\bi}{\bibitem}
\newcommand{\pav}{PAV\v SI\v C}

\begin{document}

\baselineskip 0.8cm

\vspace{1cm}

\begin{center}{\Large \bf The Classical and Quantum Theory of
Relativistic p-Branes without Constraints$^{*}$}
\footnotetext{$^{*}$Work suppoerted by CNPq, Brazil, and by the
Slovenian Ministry of Science and Technology}

\vspace{1cm}

M.Pav\v si\v c$^{**}$
\footnotetext{$^{**}$On leave from J.Stefan Institute, University
of Ljubljana, Jamova 39, 61000 Ljubljana, Slovenia}
\\
Departamento de Matem\' atica Aplicada, IMECC, Universidade Estadual
de Campinas, 13081-970, Campinas, S.P., Brazil

\end{center}

\vspace{3cm}
\thispagestyle{empty}

{\bf Abstract}

It is shown that a relativistic (i.e. a Poincar{\' e}
invariant) theory of extended objects (called p-branes) is
not necessarily invariant under reparametrizations of corresponding
$p$-dimensional worldsheets (including worldlines for $p = 0$).
Consequnetly, no constraints among the dynamical variables are
necessary and quantization is straightforward. Additional degrees
of freedom so obtained are given a physical interpretation as
being related to membrane's elastic deformations ("wiggleness").
In particular, such a more general, unconstrained theory implies
as solutions also those p-brane states that are solutions of the
conventional theory of the Dirac-Nambu-Gotto type.

\vspace{.5cm}
PACS: 1100 General theory of fields and particles; 1117 Theories of
strings and other extended objects; 1190 Other topics in general field
and particle theory

\newpage

{\bf 1. Introduction}

Quantization of relativistic extended object such as {\it p}-
dimensional membranes (often called $p$-branes) has not yet been
satisfactory solved in general. There has been much progress in
dealing with strings ($p = 1$) and point particles ($p = 0$)
\footnote{General theory of relativistic point particles and
strings is now a standard knowledge, therefore it is difficult
to cite selected particular works among so many important
original contributions. For a review and list of references see
e.g. M.Kaku \cite{1}}, but the treatment of quantized
higher dimensional objects, with
$p > 2$, in spite of important particular results \cite{2}
is not yet completed in general. Such intricaces arise
because the $p + 1$ dimensional worldsheet swept by a $p$-brane
is invariant with respect to reparametrizations; a consequence
is the existence of constraints.

An alternative approach, which has been elaborated in the case of
point partice, is to dispense with constraint and formulate the
classical and quantum relativistic theory by assuming that all
coordinates and momenta are independent \cite{3}-\cite{6}. In {\it the
unconstrained theory} mass is not fixed but occurs as a constant
of motion and a free particle still follows a straight line
with uniform speed \cite{4}-\cite{6}. Even in the presence of an
electromagnetic
field it turns out that a solution of the constrained (conventional)
theory is also a solution of the unconstrained theory \cite{4}.

In the quantized unconstrained theory the parameter $\tau$ of
evolution is explicitly present. Therefore the theory is also
called {\it the parametrized relativistic quantum mechanics}.
This elegant theory (manifestly covariant under Poincar\' e
group at every step) has been initialized by Fock and followed
by many workers \cite{3}. It is more general than the conventional
(constrained) quantum theory, since mass is not definite. But
in particular, the theory admits also the existence of definite
mass eigenstates.

In the present paper I propose to extend the unconstrained theory
of a point particle to extended objects. For this purpose I
first reformulate the constrained classical theory of a $p$-brane
by using the generalized Howe-Tucker action \cite{7} in which I isolate
$d = p + 1$ independent Lagrange multipliers by splitting the
metric tensor ${\gamma}_{ab}$ in the ADM-like manner \cite{8}. So we
obtain an action and a Hamiltonian which look like those of a point
particle \cite{9}, except for the integration over a space-like hypersurface
on the worldsheet. This reformulation of the classical constrained
$p$-brane theory is interesting in itself and possibly important for
quantization even without recourse to the unconstrained theory
which is given in Sec.4. In Sec.5 I discuss $p$-branes with
variable tension (wiggly membranes) and in Sec.6 I compare them
with the unconstrained membranes.

\vspace{1.5cm}

{\bf 2. The unconstrained point particle theory}

The idea that space-time coordinates $x^{\mu}$ of a relativistic
point particle should be considered as independent has been
pursued by many authors \cite{3}-\cite{6}. Formally this has been achieved
\cite{4}-\cite{6} by replacing the first order action
(called also phase space action) where $\lambda$ is a Lagrange multiplier
\be
   I[x,p,\lambda] = \int \dd \tau \ \left (p_{\mu} {\dot x}^{\mu}
   - {{\lambda} \over 2} (p^2 - m^2) \right )
\label{2.1}
\ee

\nnn with another action, similar in form but different in content,
\be
   I[x,p] = \int \dd \tau \ \left (p_{\mu} {\dot x}^{\mu}
   - {{\Lambda} \over 2} (p^2 - m^2) \right )
\label{2.2}
\ee

\nnn in which $\Lambda$ is not a quantity to be varied, but it is a
fixed function of the evolution parameter $\tau$. The latter action
is not invariant with respect to reparametrizations of $\tau$, therefore
there is no constraint, and all $x^{\mu}$ and $p_{\mu}$ are independent
dynamical variables. And yet (\ref{2.2}) and all equations derived
from it are invariant under Poincar\' e transformations.

Analogous procedure can be used in the second order (Howe-Tucker)
action
\begin{equation}
  I[x,\lambda] =  {1 \over 2} \int {\rm d} \tau \left( {{{\dot x}^{\mu}
   {\dot x}^{\nu}
   g_{\mu \nu}}
  \over {\lambda}} \; + \; \lambda m^2 \right)
\label{2.3}
\end{equation}

\nnn  We can replace it with another action which is solely a
functional of $x^{\mu}$:
\begin{equation}
  I[x] =  {1 \over 2} \int {\rm d} \tau \left( {{{\dot x}^{\mu} {\dot x}^{\nu}
  g_{\mu \nu}}
  \over {\Lambda}} \; + \; \Lambda m^2 \right)
\label{2.4}
\end{equation}

where $\Lambda$ is a fixed function of $\tau$ or a constant and $g_{\mu \nu}$
the metric tensor of spacetime.

The equation of motion derived from (\ref{2.4}) is
\be
   {{\dd} \over {\dd \tau}} \ \left ( {{{\dot x}_{\mu}} \over {\Lambda}}
     \right )
    - {1 \over {2 \Lambda}} \, g_{\alpha \beta , \mu}
    \ {\dot x}^{\alpha} {\dot x}^{\beta} = 0
\label{2.5}
\ee

\nnn This can be recast into a more familiar geodetic-like equation
\footnote{Eq.(\ref{2.5a}) follows directly from eq.(\ref{2.5}) if we insert
${\dot x}_{\mu} = g_{\mu \nu} {\dot x}^{\nu}$ into (\ref{2.5}) and use
${{\dd} \over {\dd \tau}} \left ( {{g_{\mu \nu} {\dot x}^{\nu}} \over
{\Lambda}} \right ) = g_{\mu \nu} {{\dd} \over {\dd \tau}} \left (
{{{\dot x}^{\nu}} \over {\Lambda}} \right )
 + g_{\mu \nu , \alpha} {{{\dot x}^{\nu}
{\dot x}^{\alpha}} \over {\Lambda}}$. The equation so obtained must then be
multiplied by $g^{\mu \rho}$ (and summed over $\mu$) and the definition
of the affinity ${\Gamma}_{\alpha \beta}^{\mu} = {1 \over 2} g^{\mu \nu}
(g_{\nu \alpha , \beta} + g_{\nu \beta , \alpha} - g_{\alpha \beta , \nu} )$
has to be taken into account.}
\be
    {1 \over {\Lambda}} {{\dd} \over {\dd \tau}} \
    \left ( {{{\dot x}^{\mu}} \over {\Lambda}} \right )
    + {\Gamma}_{\alpha \beta}^{\mu} \
    {{{\dot x}^{\alpha} {\dot x}^{\beta}} \over {{\Lambda}^2}} = 0
\label{2.5a}
\ee

\nnn where ${\Gamma}_{\alpha \beta}^{\mu}$ is the affinity composed
of the spacetime metric $g_{\mu \nu}$. In eq.(\ref{2.5}) and (\ref{2.5a})
all the variables $x^{\mu}$ are independent.

Let us now consider the quadratic form
\be
  M^2(\tau) = g_{\alpha \beta} {{{\dot x}^{\alpha} {\dot x}^{\beta}} \over
    {{\Lambda}^2}}
\label{2.6}
\ee

\nnn and calculate its derivativ with respect to $\tau$. We find

\be
    M {\dot M} = {1 \over 2} {{\dd} \over {\dd \tau}} \left (
    g_{\alpha \beta} {{{\dot x}^{\alpha} {\dot x}^{\beta}} \over
    {{\Lambda}^2}} \right ) = g_{\alpha \beta} {{{\dot x}^{\alpha}} \over
    {\Lambda}} {{\dd} \over {\dd \tau}} \ \left ( {{{\dot x}^{\beta}} \over
    {\Lambda}}
    \right ) + {1 \over 2} g_{\alpha \beta , \mu}
    {{{\dot x}^{\mu}{\dot x}^{\alpha} {\dot x}^{\beta}} \over {{\Lambda}^2}}
    = {\dot x}_{\mu} \left (
      {1 \over {\Lambda}} {{\dd} \over {\dd \tau}} \
    \left ( {{{\dot x}_{\mu}} \over {\Lambda}} \right )
    + {\Gamma}_{\alpha \beta}^{\mu} \
    {{{\dot x}^{\alpha} {\dot x}^{\beta}} \over {{\Lambda}^2}} \right ) = 0
\label{2.7}
\ee

\nnn In equating the above expression (\ref{2.7}) to zero we have used
the equation of motion (\ref{2.5a}). From eq.(\ref{2.7}) we conclude
that
\be
    M^2 = g_{\mu \nu} \ {{{\dot x}^{\mu} {\dot x}^{\nu}} \over {{\Lambda}^2}}
    = g_{\mu \nu} p^{\mu} p^{\nu} = constant
\label{2.8}
\ee

\nnn where $p_{\mu} = \p L / \p {\dot x}^{\mu} = {\dot x}^{\mu}/{\Lambda}$
is the canonical momentum. $M^2$ is thus a constant of motion even in
the presence of the background gravitational field. We may call $M$
{\it mass}, but mass is here not a fixed constant (entering the
Lagrangian, like in the conventional theory); it is an arbitrary constant
of motion, and there is no constraint among the momenta $p_{\mu}$.

By expressing $\Lambda$ in terms of $M$ and ${\dot x}^2$ (see eq.
(\ref{2.6})) we find that eq.(\ref{2.5a}) becomes indistinguishable
from the usual geodetic equation of the constraint theory:
\be
    {1 \over {\sqrt{{\dot x}^2}}} {{\dd} \over {\dd \tau}} \
    \left ( {{{\dot x}_{\mu}} \over {\sqrt{{\dot x}^2}}} \right ) +
    {\Gamma}_{\alpha \beta}^{\mu} \
    {{{\dot x}^{\alpha} {\dot x}^{\beta}} \over {{\dot x}^2}}  = 0
\label{2.9}
\ee

\nnn The last equation is reparametrization invariant, because we used
the equation (\ref{2.8}) for a fixed value of the constant of motion $M$
and because for a fixed $M$ eq.(\ref{2.8}) acts as a constraint. But
the original equation (\ref{2.5}) is not reparametrization invariant.
For a constant $\Lambda$ eq.(\ref{2.5}) has the same form as the geodetic
equation expressed in terms of proper time. The trajectory of spatial
coordinates $x^r , r = 1,2,...,D$ is the same in both the constrained and
uncontrained theory. But in the uncostrained theory the zero component,
$\mu = 0$, of equation (\ref{2.5}) has also a dynamical meaning, it is not
a redundant equation. In a previous paper \cite{4} I proposed a physical
meaning
of {\it coordinate time} $x^0$ evolving in terms of the evolution or
historical time $\tau$. According to that interpratation this expresses
the fact that an observer doesn't perceive a worldline all at once, but
instead he perceives it point by point (i.e. event by event) along the
increasing $x^0 \equiv t$. It is indeed true that in the way we perceive
the world there is something more than in the way the conventional
relativity describes it. I can only perceive the events close to
the intersection point of a time-like hypersurface (time slice or
simultaneity hypersurface)
with the worldline of my body. That is, I perceive my "now", but I
cannot perceive past or future events. And in order to be able to denote
this momentary time slice being perceived right now we need an additional
parameter besides the coordinate time $x^0$. The additional parameter is
is just $\tau$, the evolution or historical time. An observer then inferes
that the time
slice intersects also other worldlines besides his own one and that it
is moving forward in space time,
the intersection points (events) progressing along worldlines. Only
the progression of events and not the whole worldline is perceived.
The relation
$x^0 = x^0(\tau)$ traces such a progression of events on a worldline
as it is perceived by an observer.

The above interpretation obtaines an even more transparent meaning in the
quantized theory. First of all, the parametrized relativistic first and
second quantized theory is very elegant \cite{3}-\cite{6} .
Hamiltonian is not zero and it generates the true evolution which is
governed by the Schr{\"o}dinger equation:
\begin{equation}
  i {{\p \psi} \over {\p \tau}} = H \psi \, ,
  \; \; \; \; \; H = {{\Lambda} \over 2} \left ( {(-i)}^2 {\p}_{\mu}
   {\p}^{\mu} - m^2 \right )
\label{2.10}
\end{equation}

\nnn A general solution of eq.(\ref{2.10}) is given by
\be
  \psi(\tau,x) = \int {\dd}^D p \, \, c(p) e^{i p_{\mu} x^{\mu} -
  {{i \Lambda} \over 2} (p^2 - m^2) \tau}
\label{2.11}
\ee

\nnn and is normalized in {\it spacetime}:
\begin{equation}
  \int {\psi}^* (\tau,x) {\psi(\tau,x)} \, {\rm d}^4 x = 1
\label{2.11a}
\end{equation}

It may represent a wave packet which is localized in spacetime and which
moves in spacetime. The probability of finding a particle (or better an
event) at a given value of $\tau$ is different from zero in a certain
region $\Omega$ of spacetime and negligibly small (or zero) outside
$\Omega$. At later value of $\tau$ the wave packet is shifted into another
spacetime region. Thus a wave packet center sweeps a worldline in
spacetime. But at every particular value of $\tau$ a particle (event) is
most likely to be found within a particular region of spacetime,
and thus at a particular valus of coordinate time $t = x^0$.

Localization of wave function in spacetime has been usually considered as
problematic, just because it represents an instanteneous event, and
therefore it could not have been associated with a physical particle for
which probability must be conserved and unitarity of evolution operator
assured. This is indeed the case within the conventional constrained
quantum theory without a physical evolution parameter $\tau$, since in
such a theory a wave function which is localized in spacetime is
frozen for ever within a spacetime region $\Omega$. On the contrary, in the
parametrized quantum theory wave function is not frozen, and a wave packet
moves in spacetime. If a packet moves then also the probability of
observing a particle does move; at a value of the evolution parameter
${\tau}_1$ a particle is likely to be oberved at the value of the coordinate
time $t_1$, and at a later value ${\tau}_2$ the particle is likely to be
observed at another value $t_2$. Since the wave function is normalized
in spacetime (eq.(\ref{2.11a})), one immediately finds that the
$\tau$-evolution operator $U$ which brings $\psi (\tau)$ into
$\psi (\tau ') = U \psi (\tau)$ is {\it unitary}.
A more concised and detailed explanation
of the interpretation of the parametrized quantum theory is given in
ref \cite{4}.
Both first   and second quantized   parametrized theories are straightforward
and elegant. They are more general than the conventional constrained
theories, nevertheless, they contain states with definite masses and all
other results of a conventional free field theory. Extension of such a
second quantized unconstrained theory to include interactions has not
yet been fully elaborated, but significant success has been achieved at
the first quantized level \cite{10}.

\vspace{1.5cm}

{\bf 3. The separation of true Lagrange multipliers in the conventional
(constrained) p-brane action}

As a preparation for the next section in which we describe the unconstrained
p-brane theory we are now going to elaborate an ADM-like splitting of
the metric on the worldsheet swept by a p-brane. The content of this
section is intended to be self-consistent and need not be applied to
an unconstrained theory. It might be interesting to and bring new insight
to those researchers who will keep working on constrained p-brane theories.

We start from the Howe-Tucker action generalized to a membrane of
arbitrary dimension {\it p} (p-brane):
\be
  I[X^{\mu},{\gamma}^{ab}] = {{{\kappa}_0} \over 2} \int \sqrt
  {|\gamma|} ({\gamma}^{ab} {\p}_a X^{\mu} {\p}_b X_{\mu} + 2 - d )
\label{3.1}
\ee

\nnn Besides the variables $X^{\mu} (\xi) \ , \mu = 0,1,2,...,D-1$
which denote position of a $d$-dimensional ($d=p+1$) worldsheet
$V_d$ in the embedding spacetime $V_D$ , the above action contains
also the auxilliary variables ${\gamma}^{ab}$ (with a role of
Lagrange multipliers) which have to be varied independently from
$X^{\mu}$. The worldsheet parameters are ${\xi}^a , \,  a=0,1,2,...,d-1$.

By varying (\ref{3.1}) with respect to ${\gamma}^{ab}$, we arrive at
the equation for the induced metric on a worldsheet:
\be
    {\gamma}_{ab} = {\p}_a X^{\mu} {\p}_b X_{\mu} , \; \; \;
    {\p}_a \equiv {{\p} \over {\p {\xi}^a}}
\label{3.2}
\ee

\nnn Inserting (\ref{3.2}) into (\ref{3.1}) we obtain the Dirac-Nambu-
Gotto action for a minimal  surface:
\be
   I[X^{\mu}] = {\kappa}_0 \int {\dd}^d \xi \sqrt{|f|} \; , \; \;
  f \equiv {\rm det} f_{ab} , \; f_{ab} \equiv {\p}_a X^{\mu} {\p}_b X_{\mu}
\label{3.3}
\ee

\nnn The actions (\ref{3.1}) and (\ref{3.3}) are equivalent, but for the
purpose of quantization, the form (\ref{3.1}) is more convenient.

In eq.(\ref{3.1}) ${\gamma}^{ab}$ are the Lagrange multipliers, but they
are not all independent. The number of worldsheet constraints is $d$ and the
same is the number of independent Lagrange multipliers. In order to
separate out of ${\gamma}^{ab}$ the independent multipliers we proceed
as follows. Let $\Sigma$ be a space like hypersurface on the worldsheet,
and $n^a$ the normal vector field to $\Sigma$. Then the worldseet metric
tensor can be written as
\be
  {\gamma}^{ab} = {{n^a n^b} \over n^2} +  {\bar \gamma}^{ab} \quad ,
  \quad {\gamma}_{ab} = {{n_a n_b} \over n^2} +  {\bar \gamma}_{ab}
\label{3.4}
\ee

\nnn where ${\bar \gamma}^{ab}$ is projection tensor, satisfying
\be
   {\bar \gamma}^{ab} n_b = 0 , \; \; {\bar \gamma}_{ab} n^b = 0
\label{3.4a}
\ee

\nnn It projects any vector into the hypersurface to which $n^a$ is
the normal. For instance, using (\ref{3.4}) we can introduce the
tangent derivatives
\be
   {\bar \p}_a X^{\mu} = {{\bar \gamma}_a}^{\, b} \, {\p}_b X^{\mu}
   = {{\gamma}_a}^{\, b} {\p}_b X^{\mu} -
     {{n_a n^b} \over {n^2}} {\p}_b X^{\mu}
\label{3.4b}
\ee

\nnn An arbitrary derivative ${\p}_a X^{\mu}$ is thus decomposed into
a normal and tangential part (relative to $\Sigma$):
\be
    {\p }_a X^{\mu} = n_a {\p} X^{\mu} + {\bar \p}_a X^{\mu}
\label{3.4c}
\ee

\nnn where
\be
  \p \equiv {{n^a {\p}_a X^{\mu}} \over {n^2}} , \; \; \; \; \; \; \;
  n^a {\bar \p}_a X^{\mu} = 0
\label{3.4d}
\ee

\nnn Details about using and keeping the $d$-dimensional covariant notation
as far as possible are given in ref. \cite{9} . Here I shall present a shorter
and more transparent procedure, but without the covariant notation in
$d$-dimensions.

Let us take such a coordinate system in which covariant components of
normal vectors are
\be
     n_a = (1, 0, 0, . . . , 0)
\label{3.5}
\ee

\nnn From eqs.(\ref{3.4}) and (\ref{3.5}) we have
\be
   n^2 = {\gamma}_{ab} n^a n^b = {\gamma}^{ab} n_a n_b = n^0 = {\gamma}^{00}
\label{3.6}
\ee
\be
  {\bar \gamma}^{00} = 0 \; , \; \; \;  {\bar \gamma}^{0i} = 0
\label{3.7}
\ee
\begin{eqnarray}
{\gamma}_{00} & = & {1 \over {n^0}} + {\bar \gamma}_{ij} {{n^i n^j}
  \over {{(n^0)}^2}} \label{3.8a} \\
{\gamma}_{0i} & = & - {{{\bar \gamma}_{ij} n^j} \over {n^0}} \label{3.8b} \\
{\gamma}_{ij} & = & {\bar \gamma}_{ij} \label{3.8c} \\
{\gamma}^{00} & = & n^0  \label{3.8d} \\
{\gamma}^{0i} & = & n^i  \label{3.8e} \\
{\gamma}^{ij} & = & {\bar \gamma}^{ij} + {{n^i n^j} \over {n^0}} \label{3.8f}
\; ,
\;  \;  \;  i,j = 1,2,...,p
\end{eqnarray}

\nnn The decomposition (\ref{3.4c}) then becomes
\be
   {\p}_0 X^{\mu} = \p X^{\mu} + {\bar \p}_0 X^{\mu}
\label{3.9}
\ee
\be
   {\p}_i X^{\mu} = {\bar \p}_i X^{\mu}
\label{3.10}
\ee

\nnn where
\be
  \p X^{\mu} = {\dot X}^{\mu} + {{n^i {\p}_i X^{\mu}} \over {n^0}} \; ,
  \; \; \; {\dot X}^{\mu} \equiv {\p}_0 X^{\mu} \equiv {{\p X^{\mu}}
  \over {\p {\xi}^0}}
\label{3.11}
\ee
\be
   {\bar \p}_0 X^{\mu} = - {{n^i {\p}_i X^{\mu}} \over {n^0}}
\label{3.12}
\ee

\nnn As $d$ independent Lagrange multipliers can be taken $n^a = (n^0, n^i)$.
We can now rewrite our action in terms of $n^0$ and $n^i$. We insert
(\ref{3.8d}-\ref{3.8f}) into (\ref{3.1}) and take into account that
\be
     |\gamma| = {{\bar \gamma} \over {n^0}}
\label{3.13}
\ee

\nnn where $\gamma = {\rm det} \, {\gamma}_{ab}$ is the determinant of
the worldsheet metric and ${\bar \gamma} = {\rm det} \, {\bar \gamma}_{ij}$
the determinat of the metric ${\bar \gamma}_{ij} = {\gamma}_{ij} \, , \;
i, j = 1,2,...,p$
 on the hypersurface $\Sigma$.

So our action (\ref{3.1}) after using (\ref{3.8d}-\ref{3.8f}) becomes
\be
  I[X^{\mu},n^a,{\bar \gamma}^{ij}] =
{{{\kappa}_0} \over 2} \int {\dd}^d \xi {{\sqrt{\bar \gamma}} \over
{\sqrt{n^0}}} \left ( n^0 {\dot X}^{\mu} {\dot X}_{\mu} + 2 n^i
{\dot X}^{\mu} {\p}_i X_{\mu} + ({\bar \gamma}^{ij} + {{n^i n^j} \over
{n^0}}) {\p}_i X^{\mu} {\p}_j X_{\mu} + 2 - d \right )
\label{3.14}
\ee

\nnn Variation of the latter action with respect to ${\bar \gamma}^{ij}$
gives the expression for the induced metric on the surface $\Sigma$ :
\be
  {\bar \gamma}_{ij} = {\p}_i X^{\mu} {\p}_j X_{\mu} \; , \; \; \;
  {\bar \gamma}^{ij} {\bar \gamma}_{ij} = d - 1
\label{3.15}
\ee

\nnn We can eliminate ${\bar \gamma}^{ij}$ from the action (\ref{3.14})
by using the relation (\ref{3.15}):
\be
  I[X^{\mu},n^a] = {{{\kappa}_0} \over 2} \int {\dd}^d \xi {{\sqrt
  {\bar \gamma}} \over {\sqrt{n^0}}} \left ( {1 \over {n^0}}
  (n^0 {\dot X}^{\mu} + n^i {\p}_i X^{\mu})
  (n^0 {\dot X}_{\mu} + n^i {\p}_i X_{\mu}) + 1 \right )
\label{3.16}
\ee

\nnn The latter action is a functional of the worldsheet variables
$X^{\mu}$ and $d$ independent Lagrange multipliers $n^a = (n^0,n^i)$.
Varying (\ref{3.16}) with respect to $n^0$ and $n^i$ we obtain the
worldsheet constraints:
\be
     \delta n^0 \, :  \; \; \; ({\dot X}^{\mu} + {{n^j {\p}_j X^{\mu}}
  \over {n^0}} ) {\dot X}_{\mu} = {1 \over {n^0}}
\label{3.17}
\ee
\be
     \delta n^i \, :  \; \; \; ({\dot X}^{\mu} + {{n^j {\p}_j X^{\mu}}
  \over {n^0}} ) {\p}_j X_{\mu} = 0
\label{3.18}
\ee

\nnn Using (\ref{3.11}) the constraints can be written as
\be
  \p X^{\mu} \p X_{\mu} = {1 \over {n^0}}
\label{3.19}
\ee
\be
   \p X^{\mu} {\p}_i X_{\mu} = 0
\label{3.20}
\ee

\nnn The action (\ref{3.16}) contains the expression for the normal
derivative $\p X^{\mu}$ and can be written in the form
\be
  I = {{{\kappa}_0} \over 2} \int {\rm d} \tau
  {\rm d}^p \sigma
  \sqrt {| {\bar f}|} \left( {{{\p X}^{\mu }{\p X}_{\mu}}
  \over {\lambda}} + \lambda \right) \; , \; \; \;  \lambda \equiv
  {1 \over \sqrt{n^0}}
\label{3.21}
\ee

\nnn where we have written ${\dd}^d \xi = \dd \tau \ {\dd}^p \sigma$,
since ${\xi}^a = (\tau, {\sigma}^i)$ .

So we arrived at an action which looks like the well known Howe-Tucker
action for a point particle, except for the integration over space-like
hypersurface $\Sigma$, parametrized by coordinates ${\sigma}^i,
i = 1,2,...,p$.

The equations of motion for variables $X^{\mu}$ derived from (\ref{3.16})
are exactly the equations of a minimal surface give by (\ref{3.3}).

\vspace{1.5cm}

{\bf 4. Relativistic membranes (p-branes) without constraints}

In the previous section we arrived at an action (\ref{3.16}) or (\ref
{3.21}) which is equivalent to the well known Dirac-Nambu-Gotto action
for a minimal surface. Let us now consider a {\it new} action which
has the same form as (\ref{3.16}), but now instead of the Lagrange
multipliers $n^0, n^i$ it constains fixed functions $N^0 (\tau, \sigma),
\, N^i (\tau, \sigma)$ :
\be
  I[X^{\mu},n^a] = {{{\kappa}_0} \over 2} \int \dd \tau {\dd}^p
  \sigma {{\sqrt
  {\bar \gamma}} \over {\sqrt{N^0}}} \left ( {1 \over {N^0}}
  (N^0 {\dot X}^{\mu} + N^i {\p}_i X^{\mu})
  (N^0 {\dot X}_{\mu} + N^i {\p}_i X_{\mu}) + 1 \right )
\label{4.1}
\ee

\nnn To different choices of $N^0 (\tau, \sigma), \, N^i (\tau, \sigma)$
there correspond physically different actions, describing physically
different systems. A particularly simple action we have if we take
$N^i = 0$ :
\be
  I[X^{\mu}] = {{{\kappa}_0} \over 2} \int {\rm d} \tau
  {\rm d}^p \sigma
  \sqrt {| {\bar f}|} \left( {{{\dot X}^{\mu }{\dot X}_{\mu}}
  \over {\Lambda}} + \Lambda \right) \; , \; \; \;
  \Lambda \equiv {1 \over {\sqrt{N^0}}}
\label{4.2}
\ee

\nnn This action describes a continuos colletcion of {\it unconstrained
point particles}, each being described by the action (\ref{2.4}).
 \footnote{A suitable redefinition of $\Lambda$ and $\tau$ brings
(\ref{2.4}) into the form $I = {m \over 2} \int \dd \tau \
( {{{\dot X}^{\mu} {\dot X}_{\mu}} \over {\Lambda}} + \Lambda)$.}
Individual particles are labeled by the indices ${\sigma}^i$ and they
all together form a fluid localized on a continuous membrane (p-brane).
Choice of labels ${\sigma}^i$ is arbitrary. Indeed, the action (\ref{4.2})
is invariant with respect to arbitrary reparametrizations of membrane
coordinates ${\sigma}^i$. The freedom of choice of a parametrization on
a given, say initial surface $V_{\Sigma}$ , is trivial and it does not
impose any local gauge group (and constraints) among the dynamical
variables $X^{\mu}$ which depend also on the evolution parameter $\tau$.
  \footnote{Analogous situation occurs in the description of
non-relativistic (Newtonian) motion of a usual 1-dimensional string or
2-dimensional
membrane in 3-dimensional space, with the ordinary time $t$ as
evolution parameter. The fact that one can arbitrarily parametrize
string or membrane does not imply any constraints in such a non-
relativistic motion.}

The action (\ref{4.1}) or (\ref{4.2}) is not equivalent to the action
of the Dirac-Nambu-Gotto $p$-dimensional membrane. In (\ref{4.1}) and
(\ref{4.2}) all components $X^{\mu} (\tau, \sigma)$ are {\it independent
dynamical variables}. They describe motion of fluid particles in
spacetime.

Initial data may be specified on any $p$-dimensional space-like surface
$V_{\Sigma}$ embedded in $D$-dimensional spacetime. They are given by
\be
    X^{\mu} (0, {\sigma}^i) \; , \; \; \; \;  {\dot X}^{\mu} (0,
     {\sigma}^i)
\label{4.3}
\ee

\nnn Once $X^{\mu} (0,\sigma)$
on a chosen initial $V_{\Sigma}$ are determined, also a parametrization
of $V_{\Sigma}$ (i.e. choice of coordinates ${\sigma}^i$) is determined.
The dynamical equations of motion (which can be straghtforwardly derived
from (\ref{4.1}) or (\ref{4.2})) then determine $X^{\mu} (\tau,\sigma)$
at arbitrary $\tau$. Had we chosen different initial
velocities ${\dot X}^{'\, \mu} (0, \sigma)$ then we would have obtained
different $X^{' \, \mu} (\tau, \sigma)$. In particular we can choose
${\dot X}^{'\, \mu} (0, \sigma)$ so that $X^{' \, \mu} (\tau, \sigma)$
describes from the matematical point of view the same manifold $V_d$ as it
is represented by $X^{\mu} (\tau,\sigma)$. But physically,
$X^{\mu} (\tau,\sigma)$ and $X^{' \, \mu} (\tau, \sigma)$ represent
motions of different objects: the first membrane is elastically deformed
in a certain way, and the second membrane is elastically deformed in
some other way.
  \footnote{Again we have the analogy with a usual non-relativistic
elastic string or membrane. It can be elastically deformed in such a
way that the manifold $V_p$ ($p=1$ or 2) remains the same, but nevertheless
a deformed object, described by ${\bf x}' (\sigma)$, is physically
different from the "original" object described by ${\bf x} (\sigma)$.
Both ${\bf x} (\sigma)$ and ${\bf x}' (\sigma)$ describe the same
manifold $V_p$, but ${\bf x}^{'} (\sigma)$ now represents positions of
an elastically deformed string or membrane.}
This illustrates that our system is a "wiggly" membrane (see Sec.5).

The canonically conjugate variables belonging to the action (\ref{4.2})
are
\be
    X^{\mu} (\sigma) \; , \; \; \;  p_{\mu} = {{\p L} \over {\p {\dot
    X}^{\mu}}} = {{{\kappa}_0 \sqrt{|{\bar f}|} {{\dot X}_{\mu}}}
    \over {\Lambda}}
\label{4.3a}
\ee

\nnn The Hamiltonian is
\be
    H = {{\Lambda} \over 2} \int {\dd}^p \sigma  \sqrt{\bar f}
    \, \left ( {\,
    {{p^{\mu} p_{\mu}} \over {|{\bar f}|}} - {\kappa}_0^2} \right )
\label{4.4}
\ee

\nnn There are $D$ independent functions $X^{\mu} (\sigma)$ and $D$
independent functions $p_{\mu} (\sigma)$, and no constraints. Therefore
the Poisson brackets and the Hamiltonian formalism can be written
down straightforwardly (according to the lines initiated e.g. in \cite{9}).

The theory can be straightforwardly quantized by considering $X^{\mu}
(\sigma), \; p_{\mu} (\sigma)$ as operators, satisfying the commutation
relations
\be
    [X^{\mu} (\sigma), \, p_{\nu} ({\sigma}')] = {{\delta}^{\mu}}_{\nu}
    \, \delta (\sigma - \sigma ')
\label{4.5}
\ee

In the representation in which operators $X^{\mu} (\sigma)$ are diagonal
the momentum operator is given by the functional derivative
\be
     p_{\mu} = -i \, {{\delta} \over {\delta X^{\mu} (\sigma)}}
\label{4.6}
\ee

\nnn A quantum state is  represented by a wave-functional $\psi [\tau,
X^{\mu} (\sigma)]$ which depends on the evolution parameter $\tau$ and
the coordinates $X^{\mu} (\sigma)$ of our unconstrained membrane.
It satisfies the functional Schr\" odinger equation

\be
      i \, {{\p \psi} \over {\p \tau}} = H  \psi
\label{4.7}
\ee

\nnn where the Hamiltonian operator $H$ is given by eq.(\ref{4.4}) in which
$p_{\mu}$ are now operators (\ref{4.6}).

The parameter $\tau$ is invariant with respect to Lorentz transformations
and general transformations of spacetime coordinates, and $H$ is also
invariant. Therefore (\ref{4.7}) is a relativistically invariant equation,
yet it implies a state evolution (and no constraints).

A general solution to eq.(\ref{4.7}) is given by
\be
  \psi [\tau, X(\sigma)] = \int {\cal D} p(\sigma) c[p(\sigma )] e^{i H \tau}
  e^{i \int p_{\mu} (\sigma) X^{\mu} (\sigma) \dd \sigma}
\label{4.8}
\ee

\nnn where $H$ is given by (\ref{4.4}) and $p_{\mu} (\sigma)$ are now
eigenvalues of the corresponding operators. A generic wave functional,
such as given in eq.(\ref{4.8}) represents a wave packet which is a
superposition of states with definite momentum $p_{\mu} (\sigma)$. It is
localized in spacetime around a p-brane and the region of localization
proceeds, with increasing $\tau$, forward along a time-like direction and
thus sweeps a ($p+1$)-dimensional worldsheet.

A wave packet is normalized according to
\be
  \int {\cal D} X(\sigma ) \, {\psi}^{*} [\tau, X^{\mu} (\sigma)] \,
  {\psi} [\tau, X^{\mu} (\sigma)] = 1
\label{4.9}
\ee

\nnn which is a straightforward extension of the corresponding
point particle
relation (\ref{2.11a}). Since (\ref{4.9}) is satisfied at any $\tau$, the
evolution operator $U$ which brings $\psi (\tau ) \rightarrow \psi (\tau ') =
U \, \psi (\tau )$ is {\it unitary}.

Expectation value of an operator $A$ is given by
\be
   <A> = \int {\cal D} X(\sigma ) \,  {\psi}^{*} [\tau, X^{\mu} (\sigma)] \,
  A \,{\psi} [\tau, X^{\mu} (\sigma)]
\label{4.10}
\ee

The amplitude for transition from a state with definite $X_1^{\mu} (\sigma )$
at ${\tau}_1$ to a state with definite $X_2^{\mu} (\sigma )$ at ${\tau}_2$
is given by the Feynman functional integral
\be
  <X_2 (\sigma), {\tau}_2|X_1 (\sigma ), {\tau}_1> = \int {\cal D} X^{\mu}
  (\tau, \sigma) \, e^{i I[X^{\mu}]}
\label{4.11}
\ee

\nnn The functions $X^{\mu} (\tau , \sigma )$ in the expression (\ref{4.11})
represent various kinematically possible motions of an elastically deformed
membrane. Since all $X^{\mu} (\tau, \sigma )$ are physically distinguishable
there is no gauge group of transformations connecting equivalent functions
$X^{\mu} (\tau, \sigma )$. Consequently, the functional integration in
(\ref{4.11}) is straightforward, and there is no need to introduce ghosts.

\vspace{1,5cm}

{\bf 5. Relativistic membranes with variable tension - wiggly membranes;
a conventional, reparametrization invariant, description}

Let us now consider a generalization of the usual Dirac-Nambu-Gotto
$p$-dimensional membranes such that tension in general is no more a
constant. Tension is admitted to vary and this is determined by the
equations of motion. A theory of wiggly strings was considered by
Hong et al. \cite{11} and they derived equations of motion
-without using an action- by writing the
spacetime stress-energy tensor and then requiring its vanishing divergence,
${T^{\mu \nu}}_{, \nu} = 0$. Here I extend the theory to an arbitrary
p-brane.

A reparametrization invarinat action for a wiggly membrane (which to
my knowledge has not yet been explicitly written down) is:
\be
  I_W = {1 \over 2} \int {\rm d}^d \xi \, \sqrt{| \gamma|} \, \lbrack
  t^{ab}{\p}_a X^{\mu}{\p}_b X_{\mu} - \epsilon + \kappa (3 - d)
  \rbrack
\label{5.1}
\ee

\nnn where
\be
  t^{ab} = (\epsilon - \kappa) u^a u^b + \kappa {\gamma}^{ab}
\label{5.2}
\ee

\nnn is the stress-energy tensor on our $d$-dimensional worldsheet,
$\kappa (\xi )$ the tension and $u^a (\xi)$ the fluid velocity satisfying
$u^a u_a = 1$. The variables $X^{\mu}$ describe position of the worldsheet
in embedding spacetime

Action (\ref{5.1}) is invariant with respect to arbitrary reparametrizations
of worldsheet coordinates ${\xi}^a$. The theory of wiggly membranes that
we are now describing is just a straightforward extension of the usual membrane
theory in which tension $\kappa$ is constant and equal to the energy
density $\epsilon$. In the latter case the expression (\ref{5.1}) reduces
to (\ref{3.1}).

If we vary (\ref{5.1}) with respect to ${\gamma}^{ab}$ then we obtain the
worldsheet constraints which imply the expression for the induced metric
\be
 {\gamma}_{ab} = {\p}_a X^{\mu}{\p}_b X_{\mu}
\label{5.2a}
\ee

\nnn The equations of motion for $X^{\mu}$ derived from (\ref{5.1}) are
\be
 {1 \over {\sqrt{|\gamma |}}} \,
{\p}_a \left( \sqrt{| \gamma |} \, t^{ab}{\p}_b X^{\mu} \right) \equiv
  D_a (t^{ab} {\p}_b X^{\mu}) = 0
\label{5.3}
\ee

\nnn where $D_a$ is covariant derivative with respect to the worldsheet metric
${\gamma}_{ab}$.

Eq.(\ref{5.3}) can be rewritten as
\be
  D_a t^{ab} \, {\p}_b X^{\mu} + t^{ab} D_a D_b X^{\mu} = 0
\label{5.4}
\ee

\nnn Contracting the latter equation by ${\p}^c X_{\mu}$ and using
the identity (which comes from the expression for the induced metric)
\be
  D_a D_b X^{\mu} \, {\p}^c X_{\mu} \equiv 0
\label{5.5}
\ee

\nnn we obtain
\be
  D_a t^{ab} = 0
\label{5.6}
\ee

\nnn Eq.(\ref{5.4}) thus becomes simply
\be
  t^{ab} D_a D_b X^{\mu} = 0
\label{5.7}
\ee

\nnn We see that the equations of motion (\ref{5.3}) imply the law of motion
(\ref{5.6}) for the fluid velicity $u^a$ and energy density $\epsilon$
besides the law of motion (\ref{5.7}) for the embedding variables $X^{\mu}$.
In order to provide a complete description of the membrane's dynamics eq.
(\ref{5.3}) must be supplemented by an equation of state
\be
  \kappa = \kappa (\epsilon )
\label{5.8}
\ee

Let us now count the number of independent equations. Because of the
identities (\ref{5.5}) there are only $D - d$ independent
equations (\ref{5.7})
besides $d$ independent equations (\ref{5.6}). Alltogether there are are
$D$ independent equation (\ref{5.6}),(\ref{5.7})
or equivalently (\ref{5.3})).
This is the same as the number of independent variables: $d-1$ independent
$u^a$, $\, D-d$ independent $X^{\mu}$, and $\kappa$ (or equivalently
$\epsilon$, since the relation (\ref{5.8}) holds).

An equation of state (\ref{5.8}) can be arbitrarily chosen. Various
equations of state hold for various kinds of wiggly membranes. In particular,
we may choose the equation of state
\be          \epsilon = \kappa   \label{5.9}     \ee

\nnn Then the stress-energy tensor obtains the simple form
\be
     t^{ab} = \kappa \, {\gamma}^{ab}
\label{5.10}
\ee

\nnn and the equation of motion (\ref{5.6}) reads
\be
  D_a (\kappa {\gamma}^{ab}) = {\gamma}^{ab}{\p}_a \kappa
  + \kappa D_a {\gamma}^{ab} = {\p}^b \kappa = 0
\label {5.11}
\ee

\nnn which implies that tension $\kappa$ must be a constant.
    \footnote{In the last step of eq.(\ref{5.11}) we used the property that
the covariant derivative of the metric tensor is zero.}
Then eq.(\ref{5.3}) or (\ref{5.7}) becomes the equation of motion for a
Dirac-Nambu-Gotto membrane (i.e. an equation of a minimal surface):
\be
  {1 \over {\sqrt{| \gamma |}}} {\p}_a (\sqrt{| \gamma |}
  {\gamma}^{ab} {\p}_b X^{\mu}) = {\gamma}^{ab} D_a D_b X^{\mu} = 0
\label{5.12}
\ee

The theory of wiggly membranes (p-branes) is just a straightforward
interesting extension of the well known theory of membranes or p-branes
(with constraints) and it contains the latter as a particular case.

\vspace{1.5cm}

{\bf 6. Comparison of an unconstrained membrane with a wiggly membrane}

In Secs. 4 and 5 we find that an unconstrained membrane has the same
number of independent variables as a wiggly membrane.
\footnote{By "wiggly membrane" from now on I meen one described in Sec.5,
and by "unconstrained membrane" one described in Sec. 4.}
Also both kinds of
membranes suffer deformations during their motion. Un unconstrained
membrane ${\cal V}_p$ can be elastically deformed which necessarily causes
the energy density on it to vary, like on a wiggly membrane. Therefore we
expect a close relationaship between both kinds of objects.

We are now going to compare the action (\ref{4.1}) with (\ref{5.1}). For
this purpose we apply to the action (\ref{5.1})
the decomposition of the worldsheet metric as done
in eqs.(\ref{3.4})-(\ref{3.13}). After a straightforward calculation we
find the following form for the action of a wiggly membrane:
\be
  I_W = {1 \over 2} \int {\rm d}^d \xi
  {{\sqrt{\bar f}} \over {\sqrt{n^2}}} \,
  \lbrack \left( (\epsilon - \kappa) u^a u^b + \kappa {{n^a n^b} \over {n^2}}
  \right)
  {\p}_a X^{\mu} {\p}_b X_{\mu} + 2 \kappa  - \epsilon \rbrack
\label{6.1}
\ee

\nnn Writing $n^a = (n^0, n^i) \, , \;  i = 1,2,...,p$
and using (\ref{3.6}), eq.(\ref{6.1}) assumes a longer form:
\begin{eqnarray}
  I_W  = {1 \over 2} \int {\rm d}^d \xi
  {{\sqrt{\bar f}} \over {\sqrt{n^0}}} \,
  & \lbrack & \left ( (\epsilon - \kappa) u^0 u^0 + \kappa n^0
  \right ) {\p}_0 X^{\mu} {\p}_0 X_{\mu}
  + 2 \left ( (\epsilon - \kappa) u^0 u^i + \kappa n^i
  \right ) {\p}_0 X^{\mu} {\p}_i X_{\mu} \nonumber  \\
  & + & \left ( (\epsilon - \kappa ) u^i u^j + \kappa {{n^i n^j} \over {n^0}}
  \right ) {\p}_i X^{\mu} {\p}_j X_{\mu} + 2 \kappa \ - \epsilon \rbrack
\label{6.2}
\end{eqnarray}

\nnn where ${\p}_0 X^{\mu} \equiv \p X^{\mu} / \p {\xi}^0 \; , \; \,
{\p}_i X^{\mu} \equiv \p X^{\mu} / \p {\xi}^i$ and ${\xi}^a = ({\xi}^0,
{\xi}^i)$. Here $n^a$ are Lagrange multipliers, and varying (\ref{6.1}) or
(\ref{6.2}) with respect to $n^a$ gives the worldsheet constraints.

Let us now take a particular choice of coordinates ${\xi}^a$, such that
the fluid velocity becomes
\be  u^a = (u^0, 0, 0, 0, ...,0)
\label{6.3}
\ee

\nnn This means that the coordinate lines ${\xi}^i = constant$ coincide with
the worldlines of the fluid particles. Then, from the spacetime point
of view, ${\p}_0 X^{\mu}$  are the tangent vectors to the fluid worldlines.
In other words, ${\p}_0 X^{\mu}$ is spacetime velocity of a fluid particle.
At this point let us recall that in the case of an unconstrained
membrane (treated in Sec.4) the quantity ${\dot X}^{\mu} \equiv \p X^{\mu} /
\p \tau$ is also velocity of a fluid particle, and the set ${\dot X}^{\mu}
(\tau, \sigma)$ for all $\tau$, $\sigma$ is a velocity field belonging to
the bundle of fluid worldlines forming the membrane.

Coordinates ${\xi}^a$ chosen so that (\ref{6.3}) is satisfied are thus
identical with the parameters $(\tau, {\sigma}^i)$ used in the description
of an unconstrained membrane, and ${\p}_0 X^{\mu}$ in eq.(\ref{6.2}) is the
same thing as ${\dot X}^{\mu}$ in eq.(\ref{4.1}). Putting $u^i = 0$ in
eq.(\ref{6.2}) we may now identify the action (\ref{4.1}) of an unconstrained
membrane with the action (\ref{6.2}) of a wiggly membrane (remembering that
both objects have the same number of independent dynamical variables). By
comparing coefficients at ${\dot X}^{\mu} {\dot X}_{\mu}$ , ${\dot X}^{\mu}
{\p}_i X_{\mu}$ and ${\p}_i X^{\mu} {\p}_j X_{\mu}$ we find the following
relations
\be
  {1 \over {\sqrt{n^0}}} \left ( (\epsilon - \kappa){(u^0)}^2 + \kappa n^0
   \right ) = {\kappa}_0 \sqrt{N^0}
\label{6.4a}
\ee
\be
  {{\kappa n^i} \over {\sqrt{n^0}}} = {{{\kappa}_0 N^i} \over {\sqrt{N^0}}}
\label{6.4b}
\ee
\be
{{\kappa n^i n^j} \over {{(n^0)}^{3/2}}} = {{{\kappa}_0 N^i N^j} \over
{{(N^0)}^{3/2}}}
\label{6.4c}
\ee
\be
  {1 \over {\sqrt{n^0}}} (2 \kappa - \epsilon) = {{{\kappa}_0} \over
{\sqrt{N^0}}}
\label{6.4d}
\ee

\nnn In addition we have also the following relation
\be
    (u^0)^2 = {\left ({1 \over {n^0}} + {\gamma}_{ij} {{n^i n^j} \over
     {(n^0)^2}} \right )}^{-1}
\label{6.5}
\ee

\nnn which comes from ${\gamma}_{ab} u^a u^b = 1$ using (\ref{3.8a}) and
(\ref{6.3}). The quantities $n^0, \, n^i$ in (\ref{6.4a})-(\ref{6.5}) are no
more arbitrary (as they were in the action (\ref{6.1}) or (\ref{6.2})), but
are fixed by the chosen parametrization ${\xi}^a = (\tau, \sigma)$ in which
eq.(\ref{6.3}) holds. If we eliminate $n^0$, $n^i$ and $u^0$ from eqs.
(\ref{6.4a})-(\ref{6.4d}) we obtain the following relation between
$\epsilon$ and $\kappa$ :
\be
{(\epsilon - \kappa)}^2 = {\kappa}^2 - {\kappa}_0^2 - {\gamma}_{ij}
{{N^i N^j} \over {N^0}} {\left ({{{\kappa}_0} \over {\kappa}} \right )}^2
({\kappa}_0^2 + (\epsilon - 2 \kappa) \kappa)
\label{6.6}
\ee

\nnn This is {\it the equation of state} that a wiggly membrane must satisfy
in order to be equivalent to an unconstrained membrane with given
$N^0$ and $N^j$.

The relation between $\epsilon$ and $\kappa$ in eq.(\ref{6.6}) is not
unique, since $\kappa$ occurs in the 4th order. Therefore we must decide
which of the 4 branches we shall tako into account. For this purpose
we consider the property of a wiggly membrane given in eq.(\ref{5.9})-
(\ref{5.11}) stating that the equation of state $\epsilon = \kappa$ implies
constant $\kappa$. Let therefore insert $\epsilon = \kappa$ into eq.(\ref
{6.6}). We obtain
\be
  ({\kappa}^2 -{\kappa}_0^2) \left ( 1 - {\gamma}_{ij}
{{N^i N^j} \over {N^0}} {\left ({{{\kappa}_0} \over {\kappa}} \right )}^2
\right ) = 0
\label{6.7}
\ee

\nnn Among 4 solutions to eq.(\ref{6.7}) there are two solutions in which
tension $\kappa$ is a constant. This solutions are
\be   \kappa = \pm \, {\kappa}_0   \label{6.8}
\ee

\nnn In order to have always positive tension we therefore choose positive
sign in eq.(\ref{6.8}). The requirement that for $\epsilon = \kappa$ it is
$\kappa = {\kappa}_0$ then selects the right equation of state among four
relations between $\epsilon$ and $\kappa$ contained in eq.(\ref{6.6}).

Since the relation (\ref{6.6}) indeed implies that, when $\epsilon = \kappa$,
the tension $\kappa$ is constant it is therefore consistent with the
requirement of the theory of wiggly membranes. On the other hand, eq.(\ref
{6.6}) is a consequence of the relations (\ref{6.4a})-(\ref{6.4d}) which
come from the identification of the two actions, namely the one of an
unconstrained membrane and the one of a wiggly membrane. We have thus
proved that any unconstrained membrane is equivalent to a wiggly membrane
for which the equation of state (\ref{6.6}) holds.

The above relations become very simple if we consider a special subclass
of unconstrained membranes given by $N^i = 0$ (see action (\ref{4.2}). Then
from (\ref{6.4b}) $n^i = 0$ and from (\ref{6.5}) ${(u^0)}^2 = n^0$ so that
eqs.(\ref{6.4a}) and (\ref{6.4d}) read
\be
  \epsilon \ \sqrt{n^0} = {\kappa}_0 \sqrt{N^0}
\label{6.9}
\ee
\be
   {1 \over {\sqrt{n^0}}} \ (2 \kappa - \epsilon ) = {{{\kappa}_0} \over
    {\sqrt{N^0}}}
\label{6.10}
\ee

\nnn The equation of state is simply
\be
  {(\epsilon - \kappa)}^2 = {\kappa}^2 - {{\kappa}_0}^2
\label{6.11}
\ee

\nnn giving $\kappa = {\kappa}_0$ when $\epsilon = \kappa$.

\vspace{1.5cm}

{\bf 7. Conlusion}

I have investigated relativistic extended objects, called membranes or
p-branes, of arbitrary dimension $p$, including point particles when $p = 0$.
When such an object moves in spacetime it sweeps a $(p+1)$-dimensional
manifold called worldsheet. In conventional approaches the properties of
such a worldsheet are considered. The variables describing position of a
worldsheet in spacetime are not independent but satisfy $p + 1$ constraints.
In the present work I take into account the fact that our observer does
not perceive the whole worldsheet but only a space-like slice on it (this
is just what we call membrane or p-brane) and the fact that such a slice is
not at rest in spacetime but it {\it moves} into a time-like direction.
The speed of such a motion is taken as a dynamical variable and all variables
describing position of a membrane in spacetime are independent, so that there
are no constraint relations among them. The classical and quantum theory
so constructed is much more straightforward - both conceptually and
technically - and easier to handle than the constrained formalisms
of conventional p-brane theories which for $p \ge 2$ become nearly
intractable because of technical obstacles. The usual Dirac-Nambu-Gotto
membranes (p-branes) belong to a subclass of solutions
to such a theory of unconstrained $p$-dimensional membranes.

Further investigations then reveal that the unconstrained membranes are
equivalent to the so called wiggly membranes which have variable tension
$\kappa$, different from energy density $\epsilon$, provided that the latter
quantities satisfy a special equation of state $\epsilon = \epsilon (\kappa )$.

My final conclusion is therefore that instead of the theory of the
Dirac-Nambu-Gotto membranes we should rather consider the theory
of unconstrained
or wiggly membranes as an appropriate candidate for a "final" physical theory.

\end{document}